# Assessing Roundabout Safety Perceptions under Heterogeneous Traffic: Socio-Demographic and Geometric Influences in Indian Urban Contexts

Abhijnan Maji[1] and Indrajit Ghosh[2]

[1] PhD Candidate, Department of Civil Engineering, Indian Institute of Technology Roorkee, Roorkee, Uttarakhand, India, 247667.
Email: a_maji@ce.iitr.ac.in

[2] Full Professor, Department of Civil Engineering, Indian Institute of Technology Roorkee, Roorkee, Uttarakhand, India, 247667.
Email: indrafce@iitr.ac.in

**Corresponding Author:** Indrajit Ghosh (indrafce@iitr.ac.in)

**ABSTRACT**

Evaluation of the safety perceptions of roundabout users is crucial for improving road safety in mixed-traffic environments. The crash- and conflict-based analyses do not incorporate the socio-demographic characteristics of the roundabout users, which can only be captured through questionnaire surveys on a larger scale. This research evaluated the relationship of roundabout safety perception with demographic factors, driving characteristics, and varying roundabout geometries using multiple correspondence analysis, cluster analysis, factor analysis, and multinomial logistic regression. The study analyzed data from 1,530 respondents across two Indian cities. The study identified three roundabout user clusters. Single-lane roundabouts were perceived as safer during entry and circulation, with a significant prominence among middle-aged users. In contrast, double- and multi-lane roundabouts presented higher perceived risks during exit maneuvers, especially among young, inexperienced, unemployed/self-employed users. Vulnerable road users reported significantly higher perceived risks, especially under suboptimal lighting conditions. Respondents with 10-20 years of driving experience, especially car users, perceived lower risk at single-lane roundabouts but acknowledged the higher risk linked to speed variations and complex maneuvers at multi-lane roundabouts. Driving experience, vehicle type, and geometric configurations were crucial in roundabout safety perception. The study highlighted the need to improve the built environment of roundabouts for vulnerable road users. The roundabout merging area was perceived as the most dangerous spot; however, exits were also perceived as dangerous for double- and multi-lane roundabouts. The findings can benefit policymakers, engineers, and urban planners by enabling them to deploy targeted safety interventions based on issues highlighted in the study.

## 1.    INTRODUCTION

Roundabouts are at-grade intersections generally considered safer than other intersections and are widely adopted as they are known to improve traffic safety and efficiency. The implementation of roundabouts at intersections has been shown to decrease the occurrence of road traffic crashes, especially those resulting in fatalities (Persaud et al., 2001; Elvik, 2003, 2017; Vujanić et al., 2016). Research conducted in different countries has demonstrated that roundabouts can significantly enhance the operational characteristics (Mehmood and Easa, 2006; Ma et al., 2013; Al-Madani, 2022; Batari and Na'iya Ibrahim, 2024; Kabanga, 2025) and facilitate traffic safety (Bie, Lo and Wong, 2008; Chen et al., 2013; Gross et al., 2013; Jensen, 2013; Batari and Na'iya Ibrahim, 2024; Maji and Ghosh, 2025a). Notwithstanding the higher level of safety, the way road users perceive the roundabouts significantly influences their driving behavior and safety results, affecting the efficiency of the roundabouts (Leonardi, Distefano and Pulvirenti, 2020; Macioszek and Kurek, 2020). Roundabouts exhibit several benefits, such as the reduction of traffic delays and a notable decrease in the number of severe traffic collisions (Ahmad, Rastogi and Chandra, 2015). In contrast, prior research investigations on roundabout safety have shown that the perception of safety regarding roundabouts differs greatly among road users and drivers (Savolainen, Kawa and Gates, 2012; Leonardi et al., 2019; Leonardi, Distefano and Pulvirenti, 2020; Macioszek and Kurek, 2020). Typically, the crash and conflict-based analyses do not incorporate the socio-demographic characteristics of the roundabout users, which can only be captured through questionnaire surveys on a larger scale. Also, crash data in low- and middle-income countries (LMICs) often face significant problems concerning their availability, accuracy, and sufficiency (Dandona et al., 2008; Maji and Ghosh, 2025b). Therefore, investigate the various perspectives of roundabout users (motorcyclists, bicyclists, e-scooter/-bike riders, and drivers of various vehicles) who regularly use roundabouts, a comprehensive offline, face-to-face questionnaire survey was conducted. Undertaking such a study is crucial in evaluating the perception of safety among users of roundabouts, particularly in LMICs such as India, where the occurrence of roundabout crashes is notably higher than in other developed countries globally. A total of 11,410 crashes occurred at Indian roundabouts in 2022, leading to 4,051 deaths and 10,636 injuries (MoRTH, 2023). The traffic characteristics at roundabouts also influence the behavior of the drivers. Mixed traffic, prevalent in most LMICs, refers to the combined presence of different vehicle categories such as cars, motorcycles, trucks, bicycles, buses, rickshaws, pedestrians, and others (Maji et al., 2022). The presence of diverse traffic, together with lane indiscipline, results in increased complexity, confusion, and difficulty in navigating roundabouts (Vinayaraj and Perumal, 2022; Maji, Ghosh and Chandra, 2024; Maji and Ghosh, 2025b).

Geometric design is one of the most crucial determinants influencing the traffic operation and safety performance of a road network. In the case of roundabouts, the design configuration directly impacts driving behavior and the perception of potential risks while driving. While roundabouts have been extensively studied for their safety benefits in urban areas, they do not always provide sufficient levels of safety for vulnerable road users (VRUs), such as bicyclists, e-scooter/-bike riders, and motorcyclists (Rodegerdts et al., 2007; Daniels et al., 2010; Sacchi, Bassani and Persaud, 2011; Savolainen, Kawa and Gates, 2012; Pilko, Mandžuka and Barić, 2017; Oyoo, 2023; Maji and Ghosh, 2025a). Moreover, several studies have also depicted that different categories of drivers and road users based on socio-demographics have different perceptions regarding roundabout safety. Therefore, in addition to geometric design elements and traffic flow characteristics, the perception of roundabout safety is also influenced by specific socio-demographic and built-environmental factors. These factors may serve as the antecedents to improper behaviors, finally transforming into hazardous traffic situations. Therefore, the purpose of this study is to enhance comprehension of specific factors that influence the perception of risk associated with roundabouts under highly mixed, non-lane-based traffic environments. Based on the mentioned concepts and findings revealed from the literature, the research questions that have been formulated are as follows:

- How do the socio-demographic characteristics of the roundabout users correlate with the perceived risk associated with roundabout usage?

- What are the unobserved factors based on the geometric, traffic, and built-environmental characteristics affecting the safety perception among roundabout users in non-lane-based mixed traffic scenarios?
- How can the roundabout users be classified based on their socio-demographic differences, and what is the differential impact of the unobserved factors on the different segments of roundabout users?

Most of the studies in which the perceived risk in the case of roundabouts has been evaluated were conducted in developed countries where proper lane discipline prevails and the traffic composition is also less heterogeneous in nature (Distefano, Leonardi and Pulvirenti, 2018; Distefano, Leonardi and Consoli, 2019; Leonardi et al., 2019; Leonardi, Distefano and Pulvirenti, 2020). The existing studies centered on right-hand-drive traffic environments having proper lane markings and signage, and a higher rate of traffic rule compliance and lesser violations. Whereas, the traffic scenario in the case of roundabouts in India is a complete contrast in terms of traffic compositions and traffic behavior. As per the knowledge of the authors, there are very limited (or no available) studies in the roundabout safety literature where the safety perception in the case of roundabouts under non-lane-based, disordered, and highly heterogeneous traffic conditions has been examined. It is imperative to conduct a study to assess the risk perception among roundabout users and answer the aforementioned research questions, knowing the alarming roundabout crash statistics in LMICs.

This study's primary objective is to comprehensively evaluate the safety perceptions among roundabout users in India by conducting a qualitative analysis that incorporates geometric configuration, built-environmental characteristics, users' behavior, vehicular characteristics, and socio-demographic aspects. The study aims to thoroughly explore the subjective perception of roundabout safety under non-lane-based urban mixed traffic scenarios prevalent in India. The unique context of roundabout safety perception under non-lane-based, highly mixed traffic environments remained underexplored. This study addresses this gap by investigating roundabout safety perceptions in India, where diverse vehicle types and driving behaviors coexist in a complex and often chaotic traffic system. The novel contribution of this research lies in its comprehensive analysis of how socio-demographic factors influence perceived risks at roundabouts in such settings, using advanced techniques like Multiple Correspondence Analysis (MCA), Cluster Analysis, Exploratory Factor Analysis (EFA), Confirmatory Factor Analysis (CFA), and Multinomial Logistic Regression (MLR). By focusing on the intricate interactions between road users' perceptions and the built environment under mixed traffic conditions, this study provides new insights into roundabout safety challenges. An investigation into the perception and awareness of roundabout safety among its users will assist policymakers in comprehending their requirements and offering constructive suggestions for future policies and strategic choices.

The present study has been organized in the following manner: Section 2 presents a review of the previous studies in the domain of roundabout safety perception, associated analytical methods, and findings of the studies. Section 3 elaborates on the methodology involved in the present study, incorporating the questionnaire design, data collection process, and analytical methods. Section 4 discusses the results and outcomes of the analyses performed. The paper concludes by providing a summary of the study, the research challenges, and possible avenues for further research in Section 5.

## 2. LITERATURE REVIEW

Research in the field of road safety demonstrates that roundabouts are highly effective in decreasing the occurrence of accidents and facilitating a smoother and uninterrupted movement of vehicles (Bie, Lo and Wong, 2008; Chen et al., 2013; Elvik, 2017). Previously published research papers by De Brabander and Vereeck (2007) and Gross et al. (2013) have already confirmed that roundabouts are superior to signalized intersections in terms of both road users' safety and operational efficiency. Hu et al. (2014) determined that roundabouts possess the inherent capacity to significantly decrease vehicle delays and improve overall traffic management.

Several studies have shown that various groups of drivers and road users hold distinct views about roundabouts (Macioszek and Kurek, 2020). Hu et al. (2014) found that certain drivers preferred traditional signalized intersections over roundabouts due to their belief that roundabouts cause confusion, hence resulting in unsafe situations. Prior research studies have demonstrated a correlation between age and the level of acceptance towards roundabouts (Sacchi, Bassani and Persaud, 2011; Pilko, Mandžuka and Barić, 2017). It was found that younger drivers exhibited a higher level of acceptance towards roundabout usage, whereas older drivers displayed a more resistant perception. Additional research revealed that older drivers were altering their routes in order to circumvent roundabouts (Sacchi, Bassani and Persaud, 2011; Pilko, Mandžuka and Barić, 2017). Nevertheless, according to the research conducted by Retting et al. (2007), there has been a constructive change in the overall perception of roundabout users about its operational advantages. This modification highlights road users' increased awareness and active involvement in advocating for roundabouts' advantages.

The study conducted by Distefano et al. (2018) aimed to identify the geometric characteristics of roundabouts that could impact the perception of driver safety during the execution of entry, circulation, and exit maneuvers within the roundabout. The findings of the study indicate that roundabout users have a preference for single lanes on circulatory roadways. Furthermore, the single entry lane holds significant importance in contributing to the perception of driver safety. It was discovered that 60% of all crashes are caused by at least one of the geometric factors (Montella, 2011). Therefore, a key component of roundabout safety is geometric design. Numerous researchers attempted to determine how roundabouts' geometric features affected safety (Maji, Ghosh and Chandra, 2024; Maji and Ghosh, 2025a). A number of studies focused on the correlation between crash rates, traffic conditions, and the geometric design of roundabouts (Daniels et al., 2010; Anjana and Anjaneyulu, 2014; Kamla, Parry and Dawson, 2016). In addition, very few studies have been conducted to investigate the perception of risk associated with roundabouts incorporating geometry (Distefano, Leonardi and Pulvirenti, 2018; Leonardi et al., 2019).

To summarize the extensive data collected using the questionnaire survey conducted on a large number of respondents, several statistical analyses were carried out. There are very limited studies specifically examining the application of dimensionality reduction methods, such as Multiple Correspondence Analysis (MCA) and Factorial Analysis (FA), in the domain of roundabout safety under highly heterogeneous traffic environments. The study conducted by Jalayer and Zhou (2016) utilized MCA to assess the impact of roadway, environmental, motorcycle, and motorcyclist-related factors on the severity and frequency of motorcycle-related crashes. MCA was also used to examine the correlation between demographic characteristics and road traffic crash involvement by Factor et al. (2010), utilizing an extensive accident database. Das and Sun (2016) used MCA to determine the complex interplay of variables contributing to Run-Off-Road fatal crashes. Leonardi et al. (2019) used MCA to establish the relationships between geometric attributes and traffic conditions of roundabouts, as well as the characteristics of young roundabout users to assess their perception. Cluster analysis has been employed in many pre-existing studies to investigate problems in transportation systems associated with the perception of risk while driving (Oltedal and Rundmo, 2007), the likelihood of cyclists being injured in road traffic crashes (Vandenbulcke et al., 2009), the attitude of drivers regarding safety on roads (Thibenda, Wedagama and Dissanayake, 2022), and roundabout safety perception (Distefano, Leonardi and Pulvirenti, 2018). Suraji and Tjahjono (2012) employed FA to examine the impact of motorcycle-related parameters on crash risks. Distefano et al. (2018) performed FA to identify unobserved factors impacting the perception of roundabout safety. Also, FA has been employed by many transportation researchers to examine driver behavior (Şimşekoğlu and Nordfjærn, 2017; Usami et al., 2017).

The extensive review of the literature revealed that the perceptions of drivers and road users have been recognized as a significant factor influencing road safety. In addition to geometric design criteria and traffic flow characteristics, various socio-demographic and built-environmental factors also impact the perception of roundabout safety. The aforementioned elements serve as precursors to improper conduct, finally transforming into hazardous traffic situations. However, a critical review revealed that the existing studies are largely concentrated in developed countries with orderly, lane-based traffic, leaving a considerable gap in understanding these perceptions within the unique context of LMICs. These regions are often characterized by non-lane-based, highly heterogeneous, and disordered traffic environments where roundabout crash rates are alarmingly high. Furthermore, previous studies related to roundabout safety perception have often focused on narrower demographic groups, such as young people or general drivers, without fully exploring the perceptions of a wide range of road users, particularly VRUs, in the more chaotic LMIC settings. Methodologically, while prior work has utilized robust statistical tools like MCA and FA, there remains an opportunity to build a more comprehensive analytical framework that not only identifies perceptual factors but also models their differential impact across distinct user segments. This study aims to fill these contextual, demographic, and methodological voids. The present study assesses how different user groups perceive roundabout safety utilizing face-to-face questionnaire survey in an under-researched LMIC environment, broadening the analysis to include built-environmental factors and employing a multi-stage framework to provide deeper and more actionable insights.

## 3.    METHODOLOGY

This section presents the methodology followed to answer the aforementioned research questions. The framework of the complete study has been presented in Figure 1. The study has been conducted following a scientifically logical sequence, comprising questionnaire design, study area selection, data collection, and data analysis to establish the findings of the study. The primary data analysis entailed grouping roundabout users with analogous demographic characteristics (using cluster analysis), followed by identifying factors derived from perceived safety levels (using factor analysis), and assessing the impact of the extracted factors on various groups of roundabout users (using regression analysis).

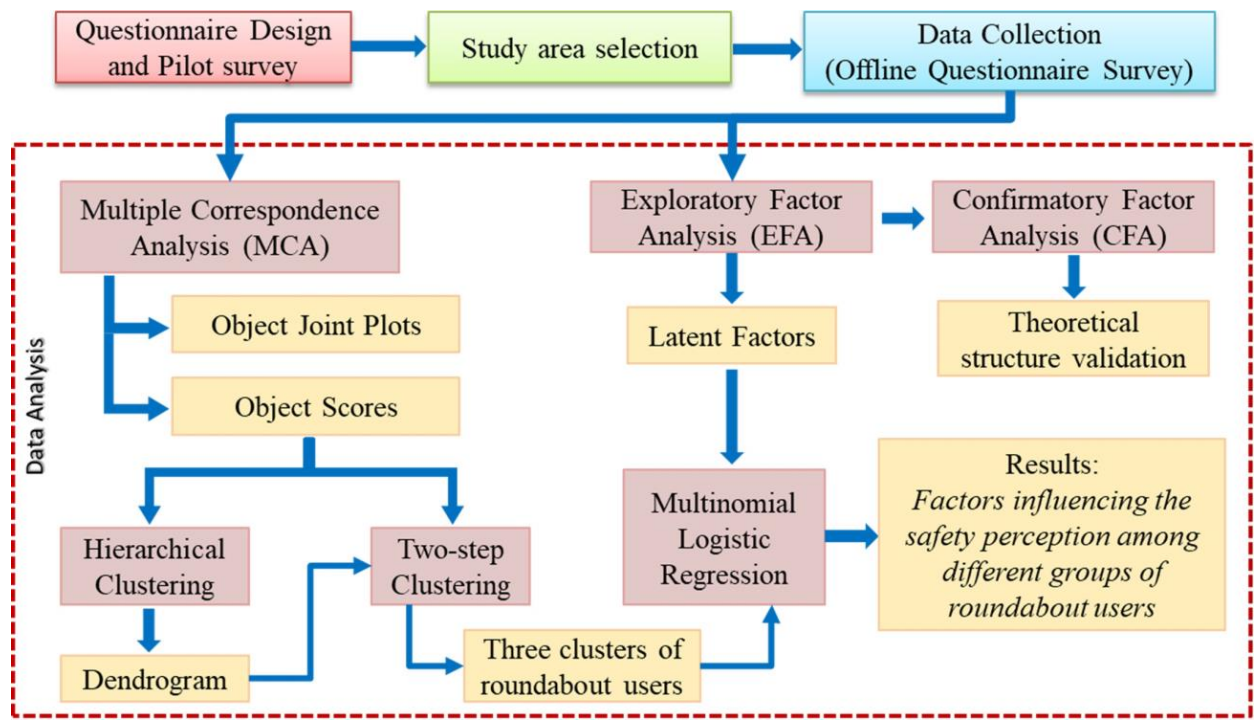

Figure 1. Overall structure of the study.

### 3.1.    Questionnaire Design

A meticulously designed questionnaire was created for an offline questionnaire survey conducted in person to evaluate the perceived levels of safety among users of roundabouts. The developed questionnaire comprised 38 items to assess the respondents' perception of roundabout safety. The survey questions were divided into four distinct sections, shown in Figure 2. The first segment gathered fundamental socio-demographic information, such as age, gender, driving experience in years, and other relevant factors. The second segment centered on the users' comprehension regarding the rules for roundabout usage. The third segment of the questionnaire survey investigated the perception of safety among individuals who use roundabout facilities like bicyclists, motorcyclists, e-scooter/-bike riders, and other vehicle drivers (i.e., from a user-specific point of view). Lastly, the fourth segment investigated the perceptions of safety at roundabouts among various roundabout users based on the roundabout's geometric characteristics (e.g., single-lane, double-lane, and multi-lane roundabouts) and various maneuver types (e.g., entering, circulating, and exiting maneuvers). The questions in the fourth segment were designed to elicit impromptu evaluations on safety, relying on the respondents' driving experiences on roundabouts without any reference to the actual roundabouts. The study also gathered overall perspectives on the perception of safety regarding different design alternatives for roundabouts. Safety perception-related questions were formulated to obtain responses on a 5-point Likert scale. The questionnaire underwent rigorous pilot studies conducted at the Indian Institute of Technology Roorkee through 15 in-person interviews conducted with professors and researchers to guarantee that the questions were skillfully designed and easily understandable to the intended respondents. The questionnaire was also translated into Hindi (i.e., the local language) to enhance understandability and usability. Again, a back-translation was conducted to identify any errors before finalizing the translation. The pilot survey served as a valuable exercise to detect problems associated with comprehension and reduce bias. The questionnaire was replicated in a Google Form to facilitate and optimize the storage of responses.

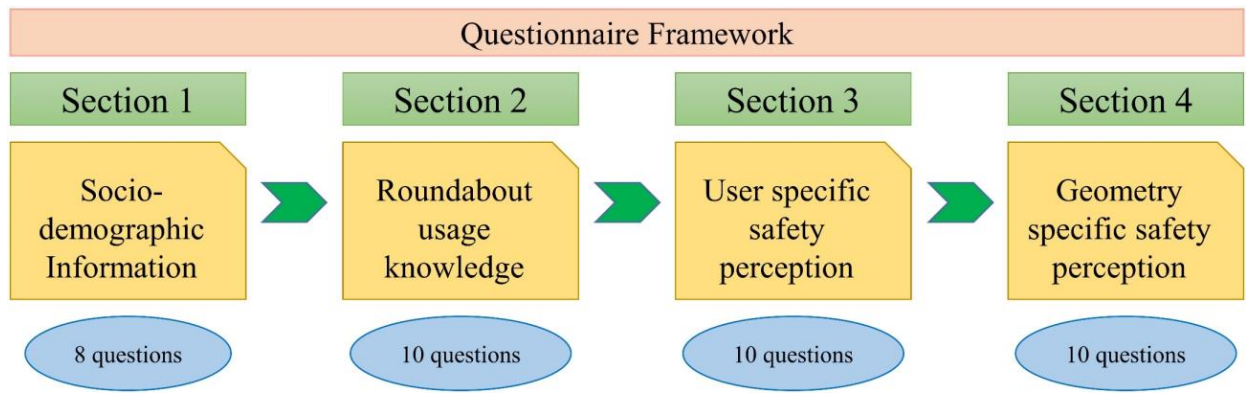

Figure 2. The overall structure of the questionnaire.

## 3.2. Study Area

The significant number of road traffic crashes in LMICs has been partially ascribed to the increase in motorized transportation, particularly in urban areas, dominated by motorcycles (MoRTH, 2019). The present study centers its attention on two cities in Northern India due to their similar degrees of motorization, the prevalence of roundabouts in the road networks, and roundabout crash statistics (MoRTH, 2023). As a case study, the cities of Chandigarh and Greater Noida were selected (shown in Figure 3). In the year 2021, there were 21 road traffic crashes at different roundabouts in Chandigarh, in which 12 people died and 20 were injured (Chandigarh Traffic Police, 2022; MoRTH, 2023). In the same year, 37 road traffic crashes occurred at different roundabouts in Greater Noida, in which 4 people died and 32 were injured (Dev et al., 2024; MoRTH, 2023). Moreover, one roundabout in Chandigarh and two roundabouts in Greater Noida were declared as black spot locations (Chandigarh Traffic Police, 2022; Dev et al., 2024). Therefore, Chandigarh and Greater Noida offer an accurate depiction of road safety concerns at the roundabouts under highly mixed, non-lane-based traffic environments observed in LMICs like India. Furthermore, the roundabouts were built more than five years prior to the initiation of the survey. This suggests that the roundabouts were not unfamiliar to the users.

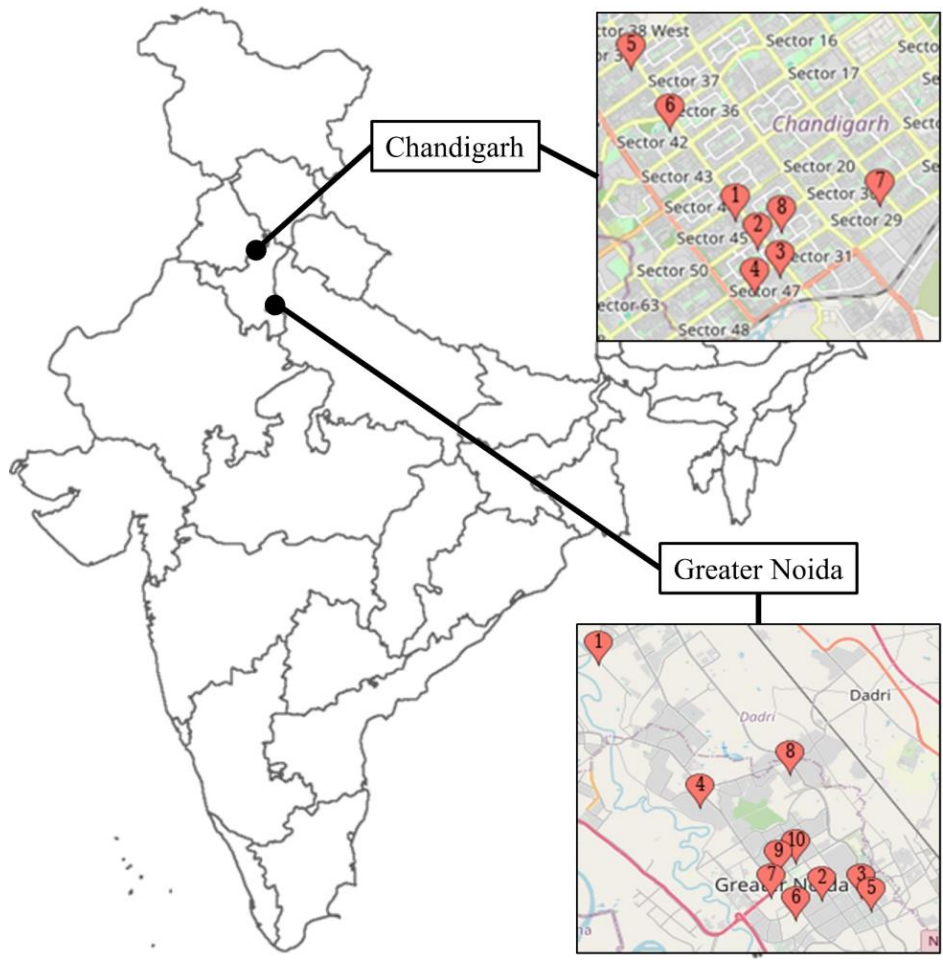

Figure 3. Location of the roundabouts in the selected cities.

### 3.3. Data Collection

The data collection process involved the recruitment and training of four graduate students from the Department of Civil Engineering, Indian Institute of Technology Roorkee. The training covered many topics, including presenting the purpose of the survey, providing an overview, and offering detailed explanations for each question. The minimum sample size required from each city was determined to be 384 at a 95% confidence level (Z = 1.96), 50% variability assumption (p = 0.5), and 5% margin of error (e = 0.05) using the Cochran's sample size calculation formula (Cochran, 1977). Data on driver population characteristics in terms of age, gender, and driving experience were unavailable for the selected two cities. Therefore, while collecting data, we aimed to assimilate data into a much larger sample size than the minimum required sample size. A larger sample size was necessary to ensure sufficient variation and a comprehensive representation of the specific variables of the population. In view of that, for the present study, we collected 782 and 748 samples from Chandigarh and Greater Noida, respectively. The collected sample size accounted for any potential loss of data due to non-responses, ensuring that the effective sample size remained adequate for meaningful statistical analyses. The population density in the southern sectors of Chandigarh (Sector 40 to 47 in Figure 3) is comparatively much higher than in other parts of the city (Chandigarh Administration, 2015). Consequently, the traffic flow at the roundabouts in these sectors is significantly higher than in the rest of the city. The bus stands, auto-rickshaw stands, car and motorcycle parking areas, shopping, and other business areas near the roundabouts in these sectors were targeted as places for data collection through offline face-to-face interviews (shown in Figure 3). Likewise, the population density and the traffic flow in the central sectors (alpha, beta, gamma, and knowledge park sectors) are substantially higher than the rest of the Greater Noida city (Greater Noida Industrial Development Authority, 2021). Therefore, the survey data were collected at the bus stands, auto-rickshaw stands, car and motorcycle parking lots, and other commercial areas near the roundabouts in the mentioned sectors (shown in Figure 3). The responses were recorded using a tablet directly into the Google form.

## 3.4.    Data Analysis

The data analyses were performed using three different software platforms: (a) Orange Data Mining tool (version 3.35), (b) IBM SPSS (version 27), and (c) SPSS Amos (version 27). This included an exploration of the connections between the socio-demographic attributes and overall safety perception regarding roundabouts among its users.

### 3.4.1.    Multiple Correspondence Analysis (MCA)

MCA is an effective tool that investigates all categories of variables in datasets (nominal, ordinal, continuous), determining the correlation among variables and providing statistical findings that are observable both analytically and visually (Costa et al., 2013; Ali et al., 2018; Thibenda, Wedagama and Dissanayake, 2022). MCA is especially pertinent in research that involves the collection of a substantial volume of qualitative data, often in conjunction with quantitative data, and where qualitative variables are susceptible to being sub-optimized during data analysis. In the present study, MCA was leveraged to graphically represent the patterns by spatially locating each item of analysis as a point in a two-dimensional space (Greenacre and Blasius, 2006). By effectively summarizing and structuring the relationships among variables, the visual illustrations offer a simplified method to comprehend and interpret large and complex datasets (Das and Sun, 2016; Ali et al., 2018; Thibenda, Wedagama and Dissanayake, 2022). The authors have utilized Orange data mining software to conduct the MCA to investigate the variations in overall risk perception at roundabouts among different respondents with driving-related characteristics (e.g., vehicle category used, frequency of roundabout usage, and driving experience) and the socio-demographic attributes (e.g., age, gender, and occupation).

### 3.4.2.    Cluster Analysis

After conducting MCA, hierarchical clustering was performed to ascertain the number of clusters utilizing the continuous object scores obtained from MCA as inputs. A dendrogram was employed to identify the optimum number of clusters in the dataset. Previous studies have utilized two-step clustering directly without using MCA and hierarchical clustering. Nevertheless, the two-step clustering method automatically establishes the number of clusters without allowing the exploration of the optimal number of clusters based on the similarity relationships among groups. Implementing hierarchical clustering allows for the determination of the optimal number of clusters by analyzing the dendrogram. The present study employed Ward's method and squared Euclidean distance to assess the similarity and distance between clusters during the hierarchical clustering process (Hair et al., 2014). Utilizing object scores from MCA as recommended by pre-existing studies (Ribbens et al., 2008) and determining the number of clusters following a two-fold clustering approach helped classify the data into multiple clusters. The reliability of the clusters was assessed using the silhouette measure of cohesion and separation, cluster size ratios (largest cluster to smallest cluster), and chi-square tests.

### 3.4.3.    Factor Analysis

A combination of Exploratory Factor Analysis (EFA) and Confirmatory Factor Analysis (CFA) approach was chosen to carry out this analysis. The primary focus of this approach is based on the theoretical constructs, which are called latent or unobserved factors. The theoretical constructs are represented by regression or path coefficients, which correspond to the relationships between the factors. CFA is a specific instance of Structural Equation Analysis, where the associations between latent variables are represented as covariance instead of explicit structural relationships, such as regressions. CFA differs from EFA in that it necessitates researchers to clearly define all aspects of the proposed measurement model (such as the number of factors and the pattern of item-factor relationships) to be investigated, whereas EFA relies more on the data.

EFA was crucial in identifying the latent and unobserved factors derived from the large initial data. The present study employed the EFA technique with the principal components approach for factor extraction and factor rotation using Varimax. The Kaiser-Meyer-Oklin (KMO) and Barlett's Tests of Sphericity were used to evaluate the appropriateness of the data for Factor Analysis. The data was considered appropriate for factorial analysis on the condition that the KMO value exceeded 0.50, and the null hypothesis (H0) was rejected with a p-value < 0.05. The criteria employed to determine the optimum number of factors to be extracted from EFA were: (a) the Kaiser criteria (i.e., factors with an Eigenvalue >1); and (ii) the Scree Plot criteria (i.e., consideration of components in the sharp descent section of the plot). Field (2009) and the manual by IBM (2017) suggested that component loading values ranging from 0.3 to 0.4 were deemed acceptable, although it was advisable to prioritize values above 0.5. In this study, the factor loadings below 0.5 were suppressed. Cronbach's alpha (α) value was computed to check the internal consistencies and to examine the reliability of the extracted factors. The desired value of Cronbach's alpha (α) lies in the range from 0.7 to 0.8 (Field, 2009). EFA was followed by the application of CFA. We were able to determine the final factorial structure utilizing the CFA, which is shown by the path diagrams. The CFA assesses the degree of fit between the hypothesized theoretical framework of the factor model and the actual data.

### 3.4.4. Multinomial Logistic Regression

The present study utilized a Multinomial Logistic Regression (MLR) model to investigate the relationship between the clusters of roundabout users (based on their demographics and driving characteristics) and their perception of roundabout safety, specifically focusing on the latent factors extracted from the EFA. The maximum likelihood approach was employed to estimate parameters by choosing coefficients that maximize the probability of observed values occurring (Field, 2009). The dependent variables in this study were the cluster results obtained from the two-step cluster analysis. The covariates were the unobserved factors derived from the EFA, which serve as the independent variables or predictors of the roundabout safety perception. During MLR analysis, it is important to consider one category as a reference for benchmarking purposes (IBM, 2017). The log-likelihood statistic, deviance statistic, and pseudo $R^2$ were employed to evaluate the model's goodness of fit by comparing the observed and predicted values of the dependent variable (Field, 2009).

## 4.    RESULTS AND DISCUSSION

## 4.1.    Survey Data Description

The summary statistics regarding the socio-demographic variables of the sample data have been presented in Table 1. In both the cities, i.e., Chandigarh and Greater Noida, it was found that the roundabouts were primarily and predominantly used by male, young, or middle-aged adult drivers who travel daily for public/private employment or education. Past research also stated that young and inexperienced drivers are more prone to experience road traffic crashes as a result of risk-taking behavior, excessive speeding, driving while distracted, and driving while under the influence of alcohol (Borowsky and Oron-Gilad, 2013; Cordellieri et al., 2016). Furthermore, the majority of drivers in the data sample used roundabouts on a daily basis and possessed driving experience ranging from 1 to 10 years. Driving experience and frequency are determining factors in the frequency and severity of crashes. Moreover, Maji et al. (2024) and Vinayaraj and Perumal (2022, 2023) unveiled that motorcycles constituted the largest proportion of motorized vehicles in India and were predominantly accountable for the occurrence of high-frequency roundabout crashes. The sample data also revealed that the majority of roundabout users utilize motorcycles (including scooters and mopeds) as their primary mode of transportation. The education level of roundabout users was also included in the data inventory as it had an impact on the frequency of road traffic crashes (Sami et al. 2013). Given that Chandigarh and Noida are located in two adjacent states in northern India, they share a similar driving culture, traffic characteristics, and roundabout user behavior. The population, demographic profile, concentration of roundabouts in the road network, and roundabout crash statistics (mentioned in Section 3.2) were also similar in both cities. Therefore, the data samples were not analyzed separately and were combined and analyzed altogether. The study included a total of 1,530 participants. The responses of the participants who did not complete the questionnaire were not submitted and stored in Google Forms. Table 2 represents the variables and their categories used in further analyses. Approximately 96% of the participants indicated that they traveled through roundabouts at least once daily. Through the analysis of the responses, it was determined that the merging area between the approaching and circular roads was perceived by 52.5% of the respondents as the most hazardous spot. Roughly 34% of the participants reported experiencing or observing crashes or near-misses while navigating the roundabouts in the study areas. The most frequent type of crash or near-miss that the participants encountered or observed was "vehicle-vehicle (front to side)" (39.3%), with the "rear-end" category following shortly thereafter (28.6%). The respondents perceived double-lane and multi-lane roundabouts to be riskier than single-lane roundabouts. The VRUs perceived higher roundabout safety concerns than the other vehicle users. Moreover, 6.6% of respondents perceived roundabouts as very safe, 34.6% as safe, 28.7% as neither safe nor dangerous, 21.5% as dangerous, and 8.6% as very dangerous.

Table 1. Socio-demographic statistics of the respondents.

|  | Category | Frequency | Proportion (%) |
|---|---|---|---|
| Gender | Female | 370 | 24 |
|  | Male | 1160 | 76 |
| Age | < 18 years | 107 | 6 |
|  | 18 to 25 years | 298 | 19 |
|  | 26 to 35 years | 840 | 55 |
|  | 36 to 50 years | 167 | 11 |
|  | > 50 years | 118 | 9 |
| Vehicle type used | Large vehicles | 107 | 7 |
|  | Cars | 321 | 21 |
|  | Auto-rickshaws | 183 | 12 |
|  | Motorcycles, E-Scooters/-Bikes | 766 | 50 |
|  | Bicycles | 153 | 10 |
| Driving experience | <1 year | 306 | 20 |
|  | 1-5 years | 648 | 43 |
|  | 6-10 years | 411 | 27 |
|  | 11-20 years | 127 | 8 |
|  | >20 years | 38 | 2 |
| Frequency of roundabout usage (/day) | 0-2 times | 169 | 11 |
|  | 3-4 times | 826 | 54 |
|  | 5-6 times | 337 | 22 |
|  | 7-8 times | 153 | 10 |
|  | > 8 times | 45 | 3 |
| Occupation | Public/Private Employee | 421 | 28 |
|  | Self Employed | 299 | 20 |
|  | Unemployed | 254 | 16 |
|  | Retired | 53 | 3 |
|  | Student | 503 | 33 |
| Education level | Primary or elementary | 66 | 4 |
|  | Middle school | 257 | 16 |
|  | High school | 485 | 32 |

| | | |
|---|---|---|
| Graduate | 597 | 39 |
| Postgraduate | 125 | 9 |

Table 2. Description of variables and their categories used in MCA and cluster analysis.

| Variables | Categories |
|---|---|
| Gender | 1: Female, 2: Male |
| Age (in years) | 1: >50, 2: 36-50, 3: 26-35, 4: 18-25, 5: <18 |
| Driving experience (in years) | 1: >20, 2: 10-20, 3: 6-10, 4: 1-5, 4: <1 |
| Vehicle Category used | 1: Large vehicle (bus, truck), 2: Car, 3: Auto-rickshaw, 4: Motorcycle, E-Scooter/-Bike, 5: Bicycle |
| Frequency of usage (times/day) | 1: 0-2, 2: 3-4, 3: 5-6, 4: 7-8, 5: >8 |
| Occupation | 1: Retired, 2: Unemployed, 3: Self-employed, 4: Public/Private Employee, 5: Student |
| Education level | 1: Primary or elementary, 2: Middle school, 3: High school, 4: Graduate, 5: Postgraduate |
| Overall safety perception | 1: Very safe, 2: Safe, 3: Neither safe nor dangerous, 4: Dangerous, 5: Very dangerous |

## 4.2. Correlation between socio-demographics and roundabout risk perception

Graphical representations provide a convenient means to efficiently condense extensive and intricate datasets to help interpret the structure of the relationships between variables and offer a unified perspective of the data (LeRoux and Rouanet, 2010). Therefore, the MCA joint plots were utilized in this study to explore and identify the relationship between various categories of socio-demographic variables and the overall roundabout risk perception. Clusters or clouds on the plots were formed by categories that have a similar distribution, while categories with different distributions were placed farther apart.

Initially, we performed the MCA utilizing three variables, i.e., age, gender, and overall safety perception. The joint plot shown in Figure 4 presents the formation of 3 point clouds. Cloud 1 represents the association of the respondent's age (between 26 and 35 years), gender (male), and overall roundabout safety perception (safe). In other words, Cloud 1 indicates that male respondents aged between 26 and 35 years perceived that roundabouts are safe. Similarly, Cloud 2 indicates that young respondents aged between 18 and 25 years perceived roundabouts as dangerous. Likewise, Cloud 3 indicates that most of the female respondents perceived roundabouts as very dangerous.

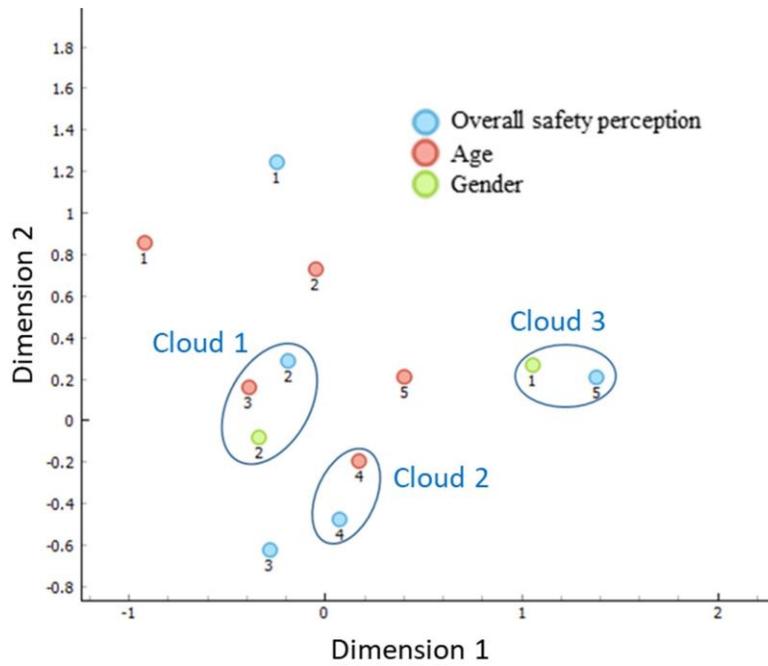

Figure 4. Combination clouds formed utilizing age, gender, and overall safety perception.

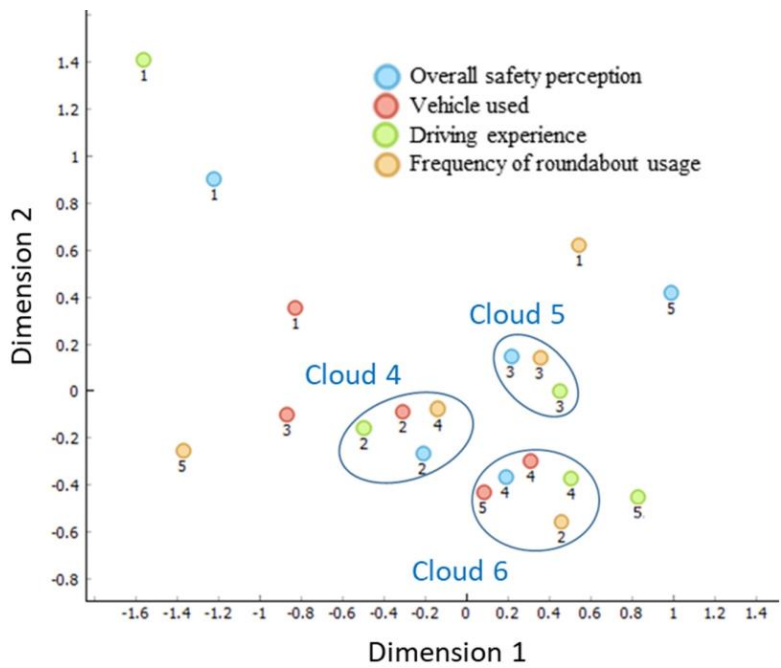

Figure 5. Combination clouds formed utilizing vehicle used, driving experience, frequency of usage and overall safety perception.

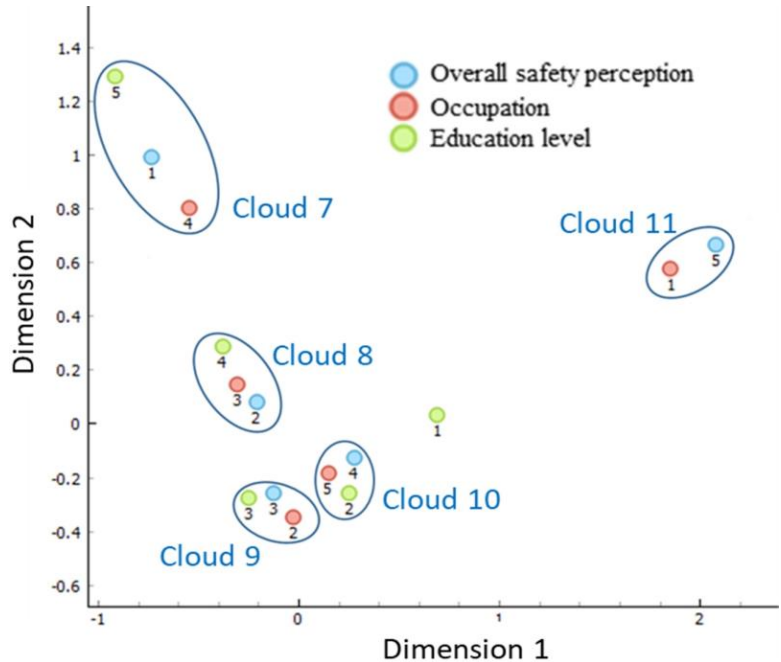

Figure 6. Combination clouds formed utilizing occupation, education level, and overall safety perception.

Subsequently, the association among vehicle category used, driving experience, frequency of roundabout usage, and overall roundabout risk perception was analyzed. 3 more point clouds were formed in the MCA joint plot, as shown in Figure 5. Cloud 4 represents the association of the vehicle type mostly used by the respondent (car), the respondent's driving experience (10 to 20 years), frequency of roundabout usage daily (7-8 times), and overall roundabout safety perception (safe). In simpler words, Cloud 4 indicates that respondents who usually drive cars and use roundabouts 7 to 8 times a day, having a driving experience of 10 to 20 years, perceived that roundabouts are safe. Similarly, Cloud 5 indicates that respondents who have a driving experience of 6 to 10 years and use roundabouts 5 to 6 times a day, perceived that roundabouts are neither safe nor dangerous. Likewise, Cloud 6 indicates that respondents who usually drive motorcycles with 1 to 5 years of driving experience and use roundabouts 3 to 4 times a day perceived roundabouts to be dangerous. Also, the respondents who preferred to bicycle perceived that roundabouts were dangerous. Finally, the association among the participants' occupation, education level, and overall risk perception was studied. Five additional point clouds were found from the MCA joint plot, as shown in Figure 6. Cloud 7 represents the association of the occupation (public or private employee), education level (postgraduate) of the participant, and overall roundabout safety perception (very safe). Therefore, Cloud 7 means that the participants who are public or private employees and received a postgraduate level education perceived roundabouts to be very safe. Cloud 8 implies that the self-employed graduate respondents perceived the roundabouts as safe. Similarly, Cloud 9 explains that the unemployed respondents who received a high school education had a mixed perception of roundabout safety (i.e., neither safe nor dangerous). However, the students who received a middle school education level perceived roundabouts as dangerous (as indicated by Cloud 10). Likewise, Cloud 11 indicates that the participants who have retired from their occupations were of the opinion that roundabouts are very dangerous.

### 4.3.    Determination of clusters within the respondent population

In this section, MCA was performed to convert categorical data into continuous data for hierarchical cluster analysis. Reportedly, the dimensions exhibited a high level of reliability ($\alpha = 0.74$) (Table 3). For dimension 1, the reliability was very high, $\alpha$ equal to 0.82. For dimension 2, the reliability was moderate, $\alpha$ equal to 0.59. The moderate reliability of dimension two can be attributed to the differential scales used (gender having two levels compared to occupation, vehicle type, and frequency of usage, which had five levels). An α value less than 0.7 has been considered acceptable in the case of exploratory research (Johnson and Wichern, 2007). The dimensions presented a significantly high percentage of variance (i.e., 77.3%) (Table 3). Hair et al. (2014) established that a variance of 95% is optimal for natural science research, while a variance of 60% or less is deemed acceptable in social sciences. MCA allowed for the visual representation of discrimination measures for each variable on every dimension (Figure 8) (Costa et al., 2013). The analysis reveals that dimension 1 exhibited stronger correlations with age, experience, and education level. Conversely, dimension 2 was linked to gender, vehicle type, occupation, and frequency of usage.

Subsequently, hierarchical cluster analysis was conducted to figure out the optimal number of clusters. This analysis was conducted using the continuous object points obtained from MCA as inputs (Figure 7). The dendrogram resulting from the hierarchical clustering technique is shown in Figure 9. The dendrogram captured three clusters (marked by the green line) of drivers that were identified from the sample data. Following that, a two-step clustering was carried out to categorize the respondents who shared similar socio-demographic traits, incorporating the object scores obtained from MCA and the optimum number of clusters derived using hierarchical clustering. High-quality clusters of the participants were generated as their silhouette measures of cohesion and separation exceeded the recommended threshold of 0.5 specified by SPSS (2011). The average silhouette score was 0.65 (shown in Figure 10). The data samples also displayed a satisfactory size ratio of the largest cluster to the smallest cluster, with a value of 2.52 (Figure 10). Moreover, chi-square tests were performed, which indicated that the variables present in each cluster had statistical significance at a 95% confidence level. Bonferroni correction was used to address the multiple comparison issues, and the significance level was set at p < 0.007 due to the inclusion of 7 demographic variables in the present study (Thibenda, Wedagama and Dissanayake, 2022). The variables include age, gender, occupation, vehicle category used, driving experience, frequency of roundabout usage, and education level.

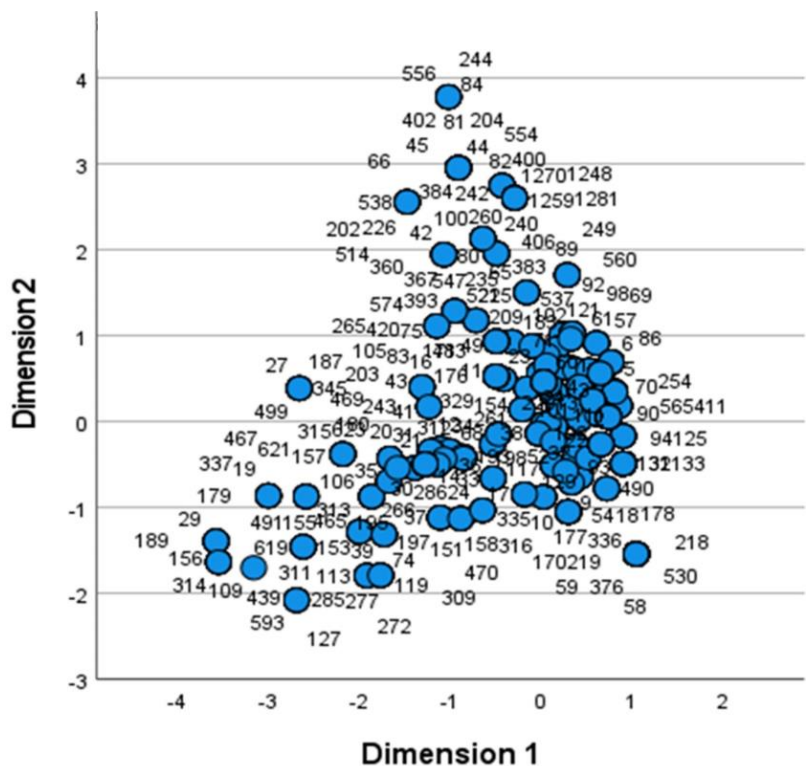

Figure 7. Object points labeled by case numbers of the variables used in MCA.

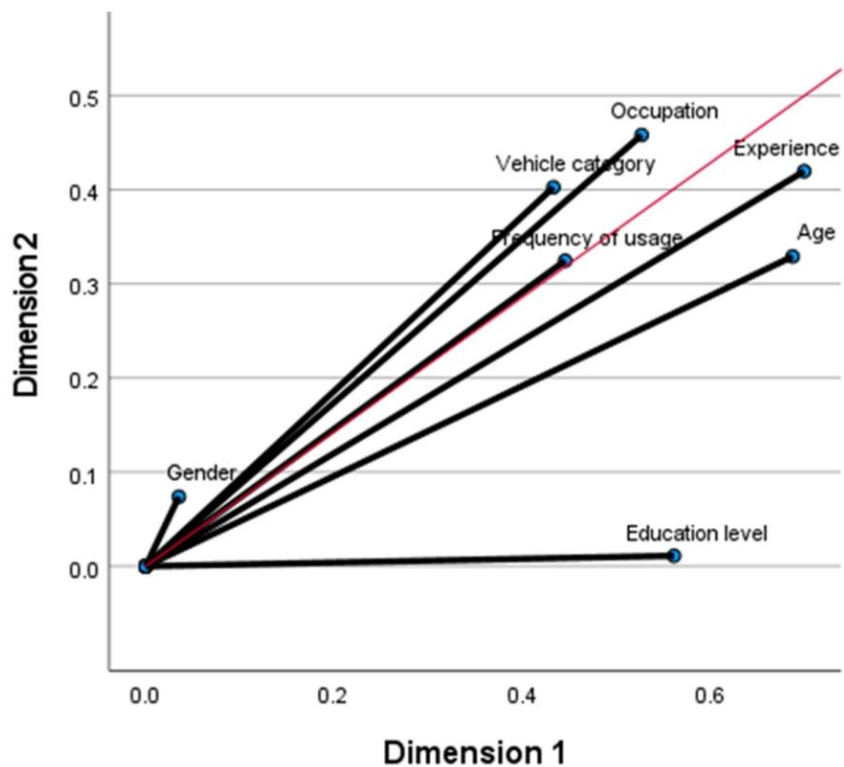

Figure 8. Discrimination measure of the variables used in MCA.

Table 3. Model summary of the variables used in MCA.

| Dimension | Cronbach's Alpha | Total (Eigenvalue) | % of Variance |
|---|---|---|---|
| 1 | 0.822 | 3.389 | 48.418 |
| 2 | 0.589 | 2.019 | 28.843 |
| Total | | 5.408 | 77.261 |
| Mean | 0.735 | 2.704 | 38.631 |

Table 4. Final outcome of the cluster analysis.

| Cluster ID | Cluster name | Frequency | Proportion (%) |
|---|---|---|---|
| Cluster 1 | Middle-aged and older male respondents with high school or graduate qualification | 811 | 53.7 |
| Cluster 2 | Middle-aged, employed, experienced males with graduate qualification | 326 | 21.3 |
| Cluster 3 | Young, unemployed, or self-employed, inexperienced users frequently using roundabouts | 383 | 25 |

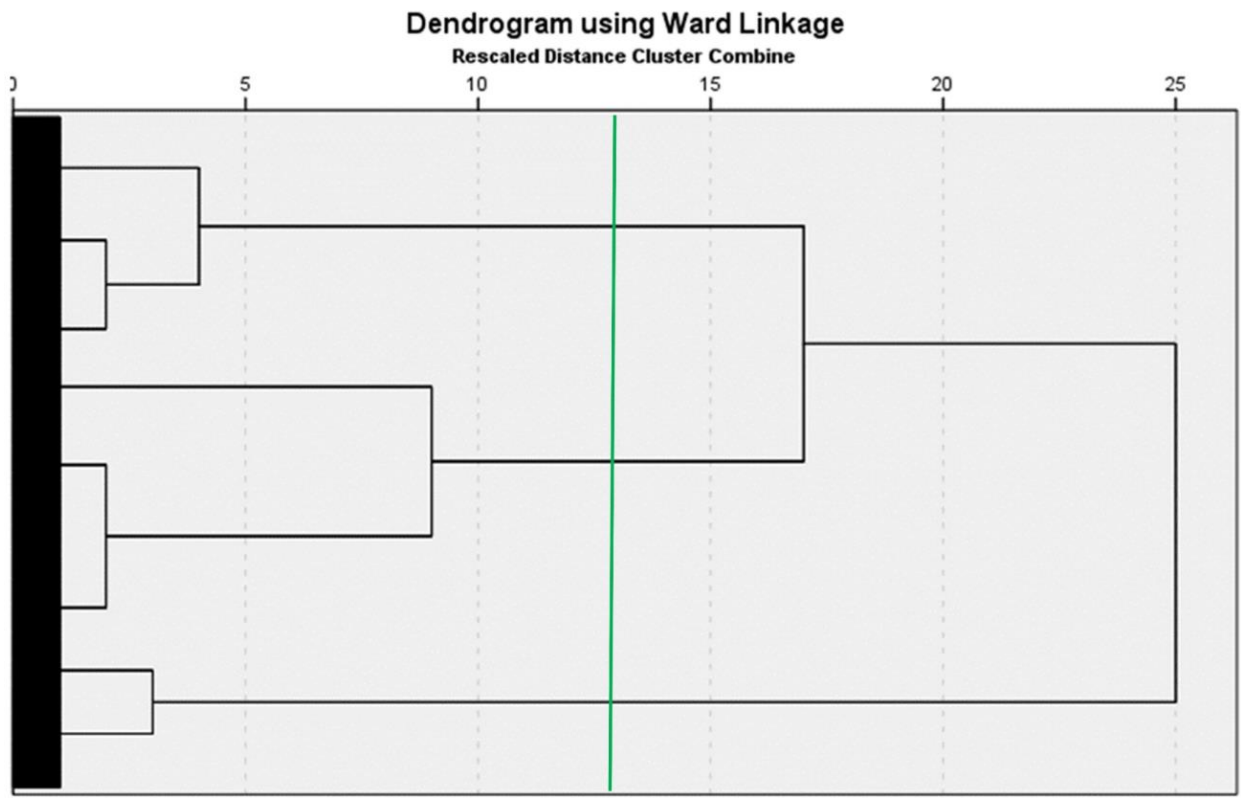

Figure 9. Dendrogram obtained from hierarchical cluster analysis.

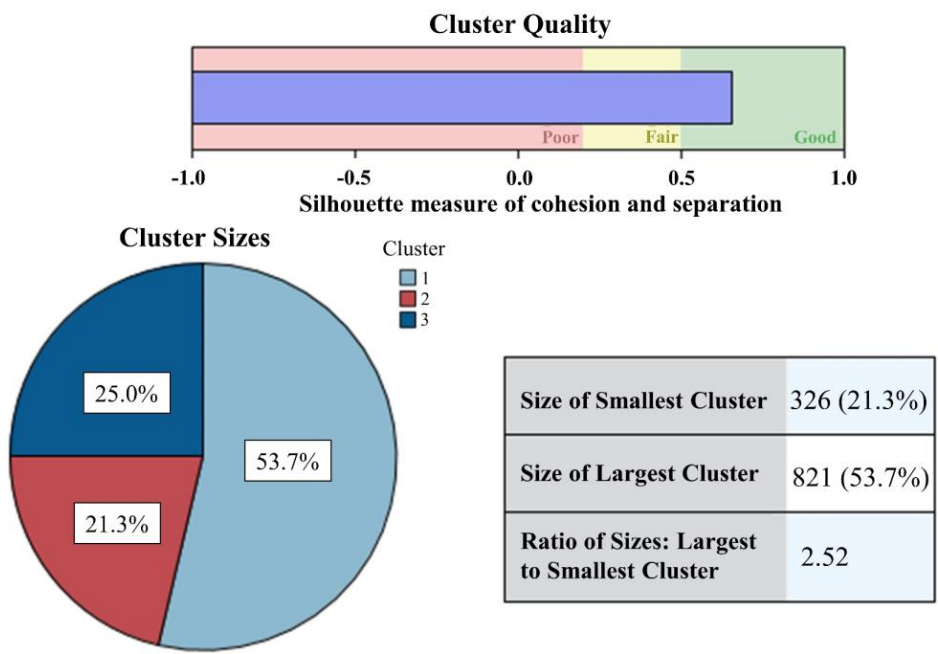

Figure 10. Cluster quality, proportion of participants in each cluster, and cluster size ratio obtained from two-step clustering.

Table 5. Description of the most frequent categories in each cluster.

| Variables | Categories | Cluster 1 Proportion (%) | Cluster 2 Proportion (%) | Cluster 3 Proportion (%) |
|---|---|---|---|---|
| Gender | Female | 23 | 14 | 32 |
| | Male | 77* | 86* | 68 |
| Age | < 18 years | 0 | 0 | 9* |
| | 18 to 25 years | 0 | 6 | 89* |
| | 26 to 35 years | 11 | 72* | 2 |
| | 36 to 50 years | 66* | 22* | 0 |
| | > 50 years | 23* | 0 | 0 |
| Vehicle type used | Large vehicles | 0 | 0 | 0 |
| | Cars | 7 | 72 | 11 |
| | Auto-rickshaws | 12 | 9 | 9 |
| | Motorcycles, E-Scooters/-Bikes | 58 | 19 | 78 |
| | Bicycles | 23 | 0 | 2 |
| Driving experience | <1 year | 7 | 0 | 21* |
| | 1-5 years | 23 | 3 | 79* |
| | 6-10 years | 32 | 13 | 0 |
| | 11-20 years | 28 | 69* | 0 |
| | >20 years | 10 | 15* | 0 |
| Frequency of roundabout usage (/day) | 0-2 times | 41 | 30 | 0 |
| | 3-4 times | 37 | 58 | 2 |
| | 5-6 times | 14 | 12 | 73* |
| | 7-8 times | 6 | 0 | 18* |
| | > 8 times | 2 | 0 | 7 |
| Occupation | Public/Private Employee | 6 | 84* | 0 |
| | Self Employed | 23 | 7 | 64* |
| | Unemployed | 26 | 4 | 31* |
| | Retired | 29 | 0 | 0 |
| | Student | 26 | 5 | 4 |
| Education level | Primary or elementary | 0 | 0 | 0 |
| | Middle school | 6 | 0 | 9 |
| | High school | 46* | 8 | 38 |
| | Graduate | 48* | 81* | 45 |
| | Postgraduate | 0 | 11* | 8 |

*Higher proportions of categories within each variable are considered when naming the clusters.

As shown in Figure 10 and Table 4, three clusters were identified in the collected sample dataset following the similar way of cluster characterization of respondents (Thibenda, Wedagama and Dissanayake, 2022). The participants were clustered into "Middle-aged and older male respondents with high school or graduate qualification," "Middle-aged, employed, experienced males with graduate qualification," and "Young, unemployed, or self-employed, inexperienced users frequently using roundabouts" based on the cluster membership of each category of variables. The specification of the participants' cluster categorizations can be found in Table 5.

The study revealed that the limited driving experience of young drivers significantly increased their likelihood of being involved in roundabout crashes. The cluster composition aligned well with the vehicular proportion statistics in India (MoRTH, 2019). The increased occurrence of road crashes among motorcyclists can be ascribed to the prevalence of motorcycles, built-in hazardous characteristics of the heterogeneous non-lane-based traffic system, and risky driving behaviors such as excessive speeding and reckless lane changing or overtaking (Tuan, 2015; Bui et al., 2020). In addition, the survey found that a considerable percentage of students ride motorcycles in the selected cities for their daily commute to school or university. Cluster 3 included the respondents falling under the age group (5–29 years) that has the highest fatality rate globally due to road traffic crashes (WHO, 2023). Borowsky and Oron-Gilad (2013) found that young, inexperienced drivers were more likely to be involved in crashes compared to experienced drivers because of their tendency to drive while distracted and exceed posted speed limits. In addition, the study also noted that a group of middle-aged male drivers commutes daily by car or motorcycle. This cluster with a relatively higher proportion of males represents the gender most susceptible to engaging in risky driving behavior. Cordellieri et al. (2016) and Oltedal and Rundmo (2007) found that the proportion of males involved in fatal crashes was double that of females. This disproportionately high rate can be attributed to the greater susceptibility of males to over-speeding, traffic law violations, and driving under the influence.

### 4.4. Determination of factors impacting roundabout safety perception

#### 4.4.1. Exploratory Factor Analysis

EFA, using the Principal Component method, was used to analyze the risk perception variables, determine and synthesize the components of users' perception towards roundabout safety. EFA is crucial in finding the connections between variables and removing redundancies (Field, 2009; Sam et al., 2019). An initial investigation examined the structure and dimensionality of the components, as well as the sufficiency of the sample size. A preliminary analysis was conducted on 20 variables. The final EFA result produced a KMO value of 0.752, greater than the acceptable threshold of 0.5. The present study considered a sample size of over 300, and the average communality values for the data sample were greater than 0.6 (refer to Table 6), the minimum threshold for applying Kaiser's criterion. An analysis of the Scree plot was carried out to ascertain the point of plateauing. A distinct descent and plateauing were observed on the Scree plot after component number 4 (shown in Figure 11). Therefore, the number of components was set at four. The four extracted components incorporated approximately 65% of the total explained variance.

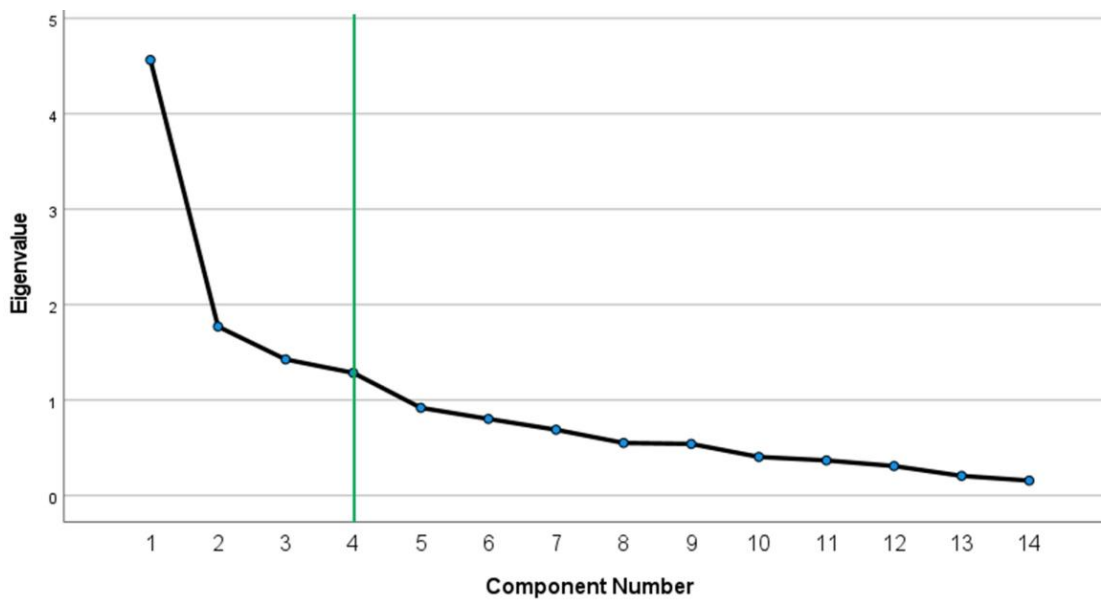

Figure 11. Scree plot obtained from EFA.

Table 6. Complete factorial structure obtained from EFA.

| Components | Items | Loading | Communality | Cronbach's Alpha (α) |
|---|---|---|---|---|
| Component 1: Risk while entering and circulating in single-lane roundabouts | Perceived safety among car drivers | 0.84 | 0.77 | 0.85 |
| | Perceived overall safety opinion | 0.75 | 0.75 | |
| | Perceived risk while entering roundabout | 0.73 | 0.64 | |
| | Perceived risk at single-lane roundabout | 0.72 | 0.60 | |
| | Perceived risk while circulating in roundabout | 0.71 | 0.62 | |
| Component 2: Risk while exiting double- and multi-lane roundabouts | Perceived Risk double-lane roundabout | 0.85 | 0.79 | 0.76 |
| | Perceived Risk multi-lane roundabout | 0.80 | 0.70 | |
| | Perceived risk while exiting roundabout | 0.73 | 0.62 | |
| Component 3: Risk due to built-environment among VRUs | Perceived safety among e-scooter/-bike riders | 0.85 | 0.75 | 0.70 |
| | Perceived safety among bicyclists | 0.79 | 0.69 | |
| | Perceived safety among motorcyclists | 0.64 | 0.68 | |
| | Perceived adequacy of road lighting | 0.57 | 0.54 | |
| Component 4: Speed change behavior while circulating and exiting roundabouts | Speed change while exiting roundabout | 0.82 | 0.70 | 0.66 |
| | Speed change while circulating roundabout | 0.81 | 0.69 | |

All component loadings below 0.5 were suppressed to facilitate better interpretation. Initially, the components were rotated using Promax and Direct Oblimin. However, the component correlation matrix indicated that the components did not exhibit any correlation. Therefore, the procedure was replicated by employing orthogonal rotation, i.e., Varimax (Brown, 2009). The first three components consisted of more than 2 variables, as shown in Table 6. Furthermore, components 1 and 2 exhibited strong reliability, as indicated by α values exceeding 0.75 (Table 6). In contrast, components 3 and 4 displayed moderate reliabilities, with α values ranging from 0.6 to 0.7. Moderate reliability may be due to weak correlations among the individual variables and a limited number of variables that contribute to the components.

The four components of the factors were named according to their consisting items as follows: (a) Component 1: "Risk while entering and circulating in single-lane roundabouts," (b) Component 2: "Risk while exiting double- and multi-lane roundabouts," (c) Component 3: "Risk due to built-environment among VRUs," (d) Component 4: "Speed change behavior while circulating and exiting roundabouts."

The derived factors indicate that the respondents perceived the risk while navigating through roundabouts because of the configuration and geometry of the roundabouts, vehicle class used, and driving behavior. Studies on roundabout safety have established that multi-lane roundabouts are more prone to unsafe situations by developing safety performance functions, crash frequency, and severity analyses (Anjana and Anjaneyulu, 2014; Distefano, Leonardi and Pulvirenti, 2018; Vinayaraj and Perumal, 2022, 2023). The present study also revealed similar results utilizing the safety perception among roundabout users. Moreover, the present study also portrayed that the roundabout users perceived the entering and exiting maneuvers as risky. To support this finding, many studies have stated that entering-circulating and exiting-circulating crashes are the type of crashes that predominantly occur in the case of roundabouts (Mandavilli, McCartt and Retting, 2009; Zheng et al., 2010; Polders et al., 2015; Maji, Ghosh and Chandra, 2024; Maji and Ghosh, 2025a). Pre-existing studies also pointed out issues regarding pedestrian and bicyclist safety at roundabouts (Rodegerdts et al., 2007; Daniels et al., 2010; Sacchi, Bassani and Persaud, 2011; Savolainen, Kawa and Gates, 2012; Pilko, Mandžuka and Barić, 2017; Maji and Ghosh, 2025a). This study also found that VRUs like bicyclists, motorcyclists, and e-scooter/-bike riders perceived roundabouts to be risky while traveling in the mixed traffic scenarios prevailing in India.

### 4.4.2. Confirmatory Factor Analysis

The factor scores from EFA were utilized to conduct CFA. The CFA provided a clear validation of the theoretical structure hypothesized using a three-factor model underlying the safety perceptions of roundabout users. The path diagram in Figure 12 shows the results of the CFA. As per Hu and Bentler (1999), the model fit indices indicated acceptable goodness of fit: Chi-square minimum divided by degrees of freedom (CMIN/DF) = 2.42, significance of Chi-square test (p-value) < 0.001; Goodness of Fit Index (GFI) = 0.95; Adjusted GFI (AGFI) = 0.91; Comparative Fit Index (CFI) = 0.95; Normalized Fit Index (NFI) = 0.94; Tucker-Lewis Index (TLI) = 0.92; Root Mean Square Error of Approximation (RMSEA) = 0.07. All factor loadings exceeded the recommended threshold of 0.5 and were statistically significant (p < 0.001) (Hair et al., 2014). Component 4 (i.e., Speed change behavior while circulating and exiting roundabouts) was removed from the CFA due to statistical insignificance. The statistical insignificance might be due to the difference in scales of the items and the number of items attributed to the factor. The values on the paths represented standardized regression coefficients, providing insights into the strength and direction of the relationships between observed variables and latent factors. In Figure 12, "Factor 1" indicates "Overall opinion and perceived risk while entering and circulating in single-lane roundabouts among car drivers", "Factor 2" indicates "Perceived risk while exiting double- and multi-lane roundabouts", and "Factor 3" indicates "Perceived risk among vulnerable road users (VRUs) and the lighting condition at roundabouts".

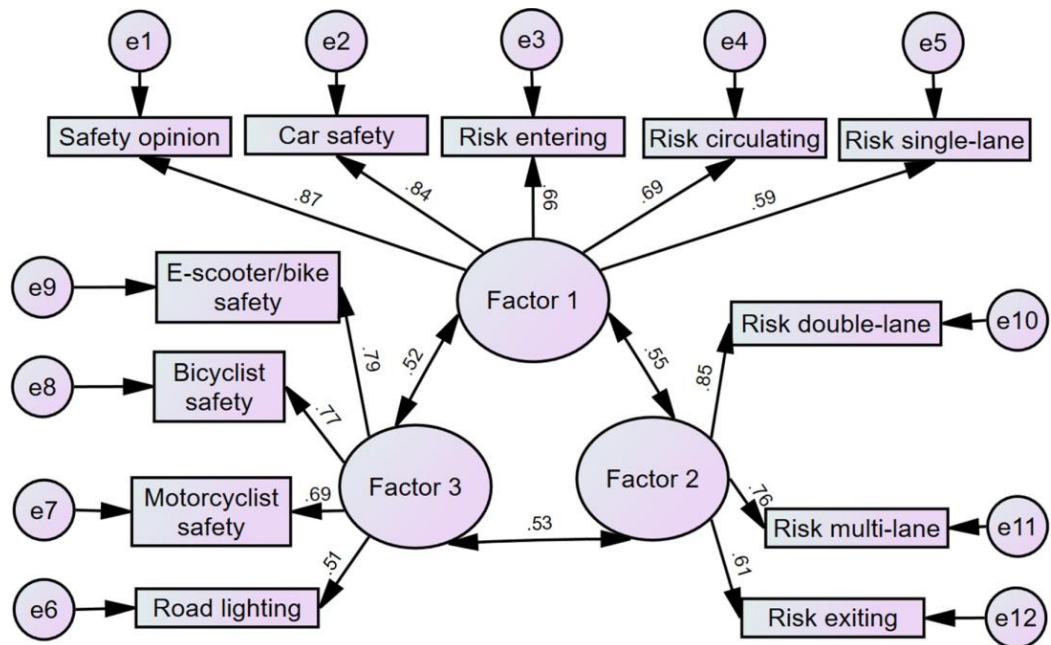

Figure 12. CFA path diagram.

Factor 1 is strongly defined by respondents' overall opinion on roundabout safety (0.87), perceived safety among car drivers (0.84), perceived risks while circulating (0.69) and entering (0.66) roundabouts, and risks perceived in the case of single-lane roundabouts (0.59). The high positive loadings across these items suggested that as drivers' concerns in these areas increase, the perceived risk associated with Factor 1 also increases significantly. In the case of Factor 2, the highest standardized regression coefficient was observed for the risk associated with double-lane roundabouts (0.85), suggesting that this was the most significant concern within this factor. The risk perception related to multi-lane configurations (0.76) and exiting roundabouts (0.61) also contributed significantly to the perception of safety in these more complex scenarios. The positive loadings across these items showed that as the perceived risks associated with exiting and navigating double- or multi-lane roundabouts increase, the overall perceived risk captured by Factor 2 also rises. Factor 3 encompassed the perceived risks faced by VRUs such as bicyclists, e-scooter/-bike riders, and motorcyclists, as well as the impact of road lighting conditions at roundabouts. The standardized regression coefficients for e-scooter/-bike riders (0.79), bicyclists (0.77), and motorcyclist safety (0.69) indicated that these were the primary concerns within Factor 3, while road lighting (0.51) was perceived as a little less critical. The positive coefficients for these items indicated that as perceived risks for VRUs and concerns about road lighting increase, the overall perceived risk captured by Factor 3 also rises substantially.

The inter-factor relationships revealed a web of interconnected perceptions where improvements in one aspect of roundabout safety could potentially influence perceptions in others. The positive relationship between Factors 1 and 2 (0.55) suggests that enhancing the entry and circulation experience in single-lane roundabouts by certain proactive safety interventions will also reduce perceived risks during double- and multi-lane roundabouts exits. The positive link between Factors 2 and 3 (0.53) indicates that addressing exit-related risks and improved lighting conditions will also benefit VRU safety at roundabouts. The relationship between Factors 3 and 1 (0.52) underscores the importance of considering the broader impact of VRU safety and built-environmental conditions on general safety perceptions regarding roundabouts. The findings of the analysis depict a multi-faceted approach to roundabout safety that will assist in suggesting safety interventions designed to address multiple factors simultaneously, leading to overall improved perceptions and actual safety outcomes.

### 4.5. Relationship between respondents' clusters and factors influencing roundabout safety

The MLR analysis was conducted to examine the relationship between the determined roundabout users' clusters and the latent or unobserved factors extracted from the EFA. The obtained results are presented in Table 7. The factors did not exhibit any correlation and clearly demonstrated that the components were mutually independent. Therefore, all the factors were utilized in developing the MLR model. The developed MLR model had Cox and Snell, Nagelkerke, and McFadden pseudo $R^2$ values of 0.43, 0.48, and 0.26, respectively, which fall within a moderate to good fit range (Ozili, 2023).

Table 7. Parameter estimate table of the multinomial logistic regression analysis.

| Clusters | Factors | B | Sig. |
|---|---|---|---|
| Cluster 1: Middle-aged and older males with high school or graduate qualification | Risk while entering and circulating in single-lane roundabouts | -1.70 | <0.001 |
| | Risk while exiting double- and multi-lane roundabouts | 0.23 | 0.004 |
| | Risk due to built-environment among VRUs | 1.27 | <0.001 |
| | Speed change behavior while circulating and exiting roundabouts | -0.30 | <0.001 |
| Cluster 2: Middle-aged, employed, experienced males with graduate qualification | Risk while entering and circulating in single-lane roundabouts | -1.28 | <0.001 |
| | Risk while exiting double- and multi-lane roundabouts | 0.15 | 0.032 |
| | Risk due to built-environment among VRUs | 0.34 | <0.001 |
| | Speed change behavior while circulating and exiting roundabouts | 0.54 | <0.001 |

Reference category: Cluster 3 (Young, unemployed or self-employed, inexperienced users frequently using roundabouts).
Cox and Snell's pseudo $R^2$: 0.43; Nagelkerke's pseudo $R^2$: 0.48; McFadden's pseudo $R^2$: 0.26

The independent variables in the model were various factors related to the respondents' opinions, perceived risks, and behaviors associated with roundabouts. The coefficients (B) represent the log-odds of being in a particular cluster compared to the reference group, and the significance levels (Sig.) indicate whether these differences are statistically significant. The model summary of the MLR analysis is presented in Table 7. The model compared two clusters (i.e., Cluster 1: middle-aged and older male respondents with high school or graduate qualifications, and Cluster 2: middle-aged, employed, experienced males with graduate qualifications) against a reference group (i.e., Cluster 3: young, unemployed, or self-employed, inexperienced users frequently using roundabouts).

The MLR model reveals that respondents in Cluster 1 (middle-aged and older male respondents with high school or graduate qualifications) are significantly less likely to perceive high risks while entering and circulating in single-lane roundabouts (B = -1.70, p < 0.001) compared to the reference group. This suggests that young, unemployed/self-employed, inexperienced users in Cluster 3 were more likely to perceive the entering and circulating maneuvers in single-lane roundabouts as risky or challenging. Conversely, Cluster 1 respondents are slightly more likely to perceive risks while exiting double- and multi-lane roundabouts (B = 0.23, p = 0.004) than those in Cluster 3, highlighting a recognition of the increased complexity of navigating exits in these roundabouts among more experienced drivers. The positive coefficient for perceived risk among VRUs and lighting conditions (B = 1.27, p < 0.001) suggests that Cluster 1 respondents are more sensitive to these safety concerns than Cluster 3, indicating that the reference group is less likely to prioritize VRU safety and adequate lighting conditions. Additionally, the negative coefficient for speed change behavior (B = -0.30, p < 0.001) indicates that Cluster 1 drivers are less likely to change their speed significantly when circulating and exiting roundabouts, suggesting that young, unemployed/self-employed, inexperienced users in Cluster 3 will exhibit more erratic driving behaviors in these situations. Similar to Cluster 1, respondents in Cluster 2 (middle-aged, employed, experienced males with graduate qualifications) are also less likely to perceive high risks while entering and circulating in single-lane roundabouts (B = -1.28, p < 0.001) compared to Cluster 3, indicating greater confidence among experienced, educated male drivers. Their perceived risk while exiting double- and multi-lane roundabouts is slightly higher (B = 0.15, p = 0.032) than that of the reference group, again suggesting an awareness of the complexities involved in these scenarios. Concerns about VRUs and lighting conditions are present at a moderate level among Cluster 2 respondents (B = 0.34, p < 0.001), but to a lesser degree than in Cluster 1. Interestingly, the positive coefficient for speed change behavior (B = 0.54, p < 0.001) in Cluster 2 suggests that these drivers are more likely to adjust their speed while navigating roundabouts compared to the reference group, indicating a more dynamic and potentially cautious driving style among experienced drivers, contrasting with the potentially more erratic or inconsistent behaviors of the young, unemployed/self-employed, inexperienced users in Cluster 3.

Previous research has also verified that the entry and the exit are crucial locations for roundabout crashes (Mandavilli, McCartt and Retting, 2009; Zheng et al., 2010; Polders et al., 2015; Novák, Ambros and Frič, 2018; Maji, Ghosh and Chandra, 2024; Maji and Ghosh, 2025a). Many studies on the safety evaluation of roundabouts stated that multi-lane and double-lane roundabouts are more susceptible to crashes than single-lane roundabouts (Anjana and Anjaneyulu, 2014; Distefano, Leonardi and Pulvirenti, 2018; Distefano, Leonardi and Consoli, 2019; Vinayaraj and Perumal, 2022, 2023; Maji, Ghosh and Chandra, 2024; Maji and Ghosh, 2025a). Moreover, a study by Leonardi et al. (2020) validates that the VRUs perceived roundabouts as more dangerous than other road users. Therefore, the existing research documents on roundabout safety support the findings of the present study. The outcomes of the present study highlight the importance of considering demographic factors when assessing safety perceptions related to roundabout use under disordered, non-lane-based, highly heterogeneous traffic scenarios. The results indicate that young, unemployed/self-employed, inexperienced users (Cluster 3) potentially engage in more erratic behaviors compared to the more confident and experienced drivers in Clusters 1 and 2. These findings underscore the need for targeted interventions that address the specific safety concerns and behaviors of different user groups to improve roundabout safety for all road users.

## 5.    CONCLUSIONS

The present study conducted an in-depth evaluation of roundabout safety perceptions among its users in two Indian cities, with a focus on investigating how socio-demographic factors influence perceived risks at roundabouts in mixed traffic conditions. By employing a combination of MCA, Cluster Analysis, EFA, CFA, and MLR analysis, the research uncovered significant patterns that differentiated the safety perceptions of various roundabout user groups, particularly in relation to roundabout design and usage. The study also highlighted that roundabouts are perceived differently by users depending on their age, driving experience, vehicle type, and daily usage patterns, despite being considered safer than other intersection types. The findings underscored the complexity of safety perceptions in environments characterized by non-lane-based, highly heterogeneous traffic–a common scenario in LMICs like India. This research contributed valuable insights that could be utilized to enhance the design and safety of roundabouts, ultimately improving road safety for all users. The major findings of this study are as follows:

- The study determined that the merging area between the approaching and circular road was perceived as the most hazardous spot by a majority of respondents (52.5%). This is corroborated by the finding that 34% of participants have directly experienced or observed crashes or near-misses. The most common of these incidents were "front to side" (39.3%) and "rear-end" crashes (28.6%), confirming that the merging areas, i.e., entering-circulating areas, are the areas of major concern. 8.6% of the respondents perceived roundabouts as very dangerous, 21.5% as dangerous, and 28.7% as neither safe nor dangerous, reflecting the antecedents to the high number of roundabout crashes in India.

- The study revealed that single-lane roundabouts were perceived as safer; however, risk perception is maneuver-dependent. While the roundabout users perceived single-lane roundabouts as safer, the study concludes that this perception is most critical during entry and circulation maneuvers. Also, such perceptions of single-lane roundabouts were pronounced among more experienced users.

- Conversely, double- and multi-lane roundabouts were associated with higher perceived risks, particularly during exiting maneuvers. This finding was quantitatively supported by higher standardized regression coefficients in the CFA for perceived risks associated with double- and multi-lane roundabouts.

- Three distinct clusters with unique perceived roundabout safety profiles were identified through hierarchical and two-step cluster analyses based on MCA object scores. These clusters were: (a) Middle-aged and older male respondents with high school or graduate qualifications, predominantly car users, who perceived roundabouts as generally safe, especially in single-lane configurations; (b) Middle-aged employed males with graduate qualifications, who showed confidence in navigating roundabouts, nonetheless, were more aware of the complexities involved in multi-lane configurations; (c) Young, unemployed or self-employed, inexperienced users, who frequently use roundabouts perceived higher risks across all types of roundabouts, especially during entry and circulation. The identification of the aforementioned clusters and their differential perceptions is a key contribution, allowing for highly targeted safety interventions.

- The study identified that VRUs, including bicyclists, e-scooter/-bike riders, and motorcyclists, perceived significantly higher risks at roundabouts. This was particularly evident in scenarios involving poor lighting conditions, where the standardized regression coefficients for VRU-related risks were notably high (e.g., 0.77 for bicyclists' safety). The CFA model indicated that improvements in lighting and infrastructure tailored to VRUs would substantially reduce these perceived risks. This concludes that a holistic approach to roundabout safety must extend beyond geometric design to include infrastructure supporting all user types.

- The analysis demonstrated that driving experience plays a crucial role in safety perception. Respondents with 10-20 years of driving experience, particularly those using cars, exhibited lower risk perceptions during single-lane roundabout maneuvers; however, they recognized the increased risks associated with speed changes and complex maneuvers in multi-lane roundabouts. The MLR results supported these, showing statistically significant differences in risk perceptions between clusters, particularly concerning speed change behaviors during circulation and exit phases.

However, there are a few limitations associated with the present study. The present study is limited by its geographic focus on two cities in Northern India, which may affect the generalizability of the findings to other regions or countries with different traffic dynamics and roundabout designs. As the participants were recruited from high-traffic locations like bus stands, parking areas, and markets near the selected roundabouts, the sample data may incorporate underrepresentation of certain demographic groups, such as older adults who may not frequent these busy commercial areas, residents from surrounding rural areas, and individuals who use roundabouts infrequently. The research was not validated using actual crash data due to the unavailability and unreliability of such data (characteristic problem associated with crash data in LMICs), especially in the case of roundabouts, which may limit the study's comprehensiveness. Future research should aim to address the limitation by expanding the geographic scope to include diverse locations, incorporating stratified or random sampling across a wider geographic area, and conducting longitudinal studies to capture evolving safety perceptions in response to infrastructural or policy changes. Integrating objective safety data, such as crash statistics or near-miss incidents, with perceived safety measures would provide a more comprehensive understanding of roundabout safety. The findings of this research have significant practical implications for policymakers, engineers, and urban planners in enhancing roundabout safety, optimizing traffic flow, and protecting VRUs. Targeted safety interventions, such as enhanced signage, speed calming devices (e.g., transverse rumble strips, speed breakers etc.), improved lighting, and better facilities for VRUs (dedicated bicycle tracks, table top bicycle and pedestrian crossings, use of rectangular rapid flashing beacons, splitter island refuges, etc.) need to be deployed at the roundabout locations based on the findings and insights from the present study to reduce perceived risks at roundabouts. Additionally, safety campaigns should be carried out to educate roundabout users and encourage safer driving behaviors, ultimately contributing to a safer driving environment at roundabouts in mixed-traffic environments.

## ACKNOWLEDGEMENT

Abhijnan Maji wants to extend his sincere thanks to the Ministry of Education (MoE), Government of India (GoI), for the Prime Minister Research Fellowship (PMRF) (ID: 2802857), which helped a lot to carry out this study.

## DECLARATION OF INTEREST STATEMENT

The authors declare that they have no known competing financial interests or personal relationships that could have appeared to influence the work reported in this paper.

## FUNDING DETAILS

This research did not receive any specific grant from funding agencies in the public, commercial, or not-for-profit sectors.